\documentclass[preprint,11pt]{JHEP3}

\JHEPspecialurl{http://jhep.sissa.it/JOURNAL/JHEP3.tar.gz}

\usepackage{epsfig,multicol,amsmath}

\usepackage{epstopdf}
\usepackage{bm}  
\usepackage{amsmath}     
\usepackage{epsfig,epsf}
\usepackage{graphicx}
\usepackage{rotating}
\usepackage{cite}

\def\beq{\begin{equation}}
\def\eeq{\end{equation}}
\def\bea{\begin{eqnarray}}
\def\eea{\end{eqnarray}}

\def\eq#1{{Eq.~(\ref{#1})}}
\def\fig#1{{Fig.~\ref{#1}}}
\newcommand{\bas}{\bar{\alpha}_S}
\newcommand{\as}{\alpha_S}

\newcommand{\Lb}{\left(}
\newcommand{\Rb}{\right)}
\parskip=\itemsep               

\newcommand{\nn}{\nonumber}

\newcommand{\h}{\frac{1}{2}}

\newcommand{\ga}{\gamma}

\title{Non-linear equation: energy conservation and  impact parameter dependence}
\author{\Large Andrey Kormilitzin${}^{a}$ \thanks{Email: andreyk1@post.tau.ac.il.}\,\,\,\,and\,\,
Eugene\, Levin${}^{a, b}$ \thanks{Email: leving@post.tau.ac.il.}\,
\\
${}^a$ \, Department of Particle Physics, School of Physics and Astronomy,
Tel Aviv University, Tel Aviv, 69978, Israel\\
${}^b$\, Departamento de F\'\i sica, Universidad T\'ecnica
Federico Santa Mar\'\i a, Avda. Espa\~na 1680,
Casilla 110-V,  Valparaiso, Chile}

\abstract{In this paper we address two questions: how  energy conservation affects the solution to the non-linear equation, and how  impact parameter dependence influences  the inclusive production.
Answering the first question
we solve the modified BK equation which takes into account  energy conservation. In spite of the fact that we used the simplified kernel, we believe that the main result of the paper: the small ($\leq  40\%$) suppression  of the inclusive production
due to energy conservation,  reflects a general feature. This result leads us to believe that the small value of the nuclear modification factor is of a non-perturbative nature.  In the solution a new scale appears $Q_{fr} = Q_s \exp(-1/(2 \bas))$  and the production of  dipoles with the size larger than $2/Q_{fr}$ is suppressed. Therefore, we can expect that the typical temperature for hadron production is about $Q_{fr}$ ($ T \approx Q_{fr}$). The simplified equation allows us to obtain a solution to Balitsky-Kovchegov equation taking into account the impact parameter dependence. We show that the impact parameter ($b$) dependence can be absorbed into the non-perturbative  $b$ dependence of the saturation  scale.  The solution of the BK equation, as well as of the modified BK equation without $b$  dependence,  is only accurate up to  $\pm 25\%$.}


\dedicated{PACS: 13.85.-t, 13.85.Hd, 11.55.-m, 11.55.Bq}

\preprint{TAUP 2920/10\\
{\tt }\\
\today}

\begin{document}

\section{Introduction}
\setcounter{equation}{0}

High density QCD that has
 a long history \cite{GLR,MUQI,MV,MUCD,BK,JIMWLK},
it  is actually  based on the idea that before summing the next-to-leading order correction to the BFKL Pomeron \cite{BFKL},
we need to deal with  different kinds of corrections. Indeed, the scattering amplitude  in the leading order BFKL approach
is proportional  to
\beq \label{NLL1}
 N_{\mbox{BFKL}}\,\,\,\propto\,\,\,\as^2 \,s^{ \Delta_{BFKL}}
\eeq
where $\Delta^{LO}_{BFKL} \propto \as$ .  We can see two kind of corrections to the amplitude of \eq{NLL1}: the first one is the next-to-leading order correction to the BFKL kernel that gives $\Delta_{BFKL} = \as C_1 + \as^2 C_2$ where $C_1$ and $C_2$ are constants; and the second corrections due  to the  exchange of  two BFKL Pomerons, namely,

\beq \label{NLL2}
N_{\mbox{2 BFKL Pomeron exchange}}\,\,\,\propto\,\, N^2_{\mbox{BFKL}}\,\,\,\propto\,\,\as^4 \,s^{2 \Delta_{BFKL}}\,
\eeq
Comparing \eq{NLL1} and \eq{NLL2} one sees that for a wide range of
 energies
\beq \label{NLL3}
y\,\,=\,\,\ln s\,\,>\,\,\frac{2}{\as}\,\ln \Lb \frac{1}{\as} \Rb
\eeq
One has also to include the exchange of two BFKL Pomerons that leads to  non-linear corrections in the evolution equations.

For these energies the NLO corrections to the BFKL kernel are small . They becomes essential for higher energies  for  which $\as^2 y \approx 1$ or $y = \ln s > 1/\as$. Therefore, for energies in the kinematic region
\beq \label{KR}
1/\as^2\, \geq \, y \, \geq \,\frac{2}{\as}\,\ln \Lb \frac{1}{\as} \Rb
\eeq
we have to include multi-Pomeron exchange and can   neglect  NLO corrections to the BFKL kernel.

Explicit calculations of the NLO corrections to the BFKL kernel
 \cite{NLL}, show that $\as^2$ corrections to $\Delta_{BFKL}$ are
 rather large,  and have to be taken into account for any realistic and practical  estimates. In terms of \eq{KR} the large NLO corrections mean that the kinematic range for the high density QCD is narrow.   Fortunately, we know the NLO corrections to the linear BFKL equations \cite{NLLG} and to the non linear (Balitsky-Kovchegov) equation  (see Ref. \cite{BAREV} for review and references) quite well.  We  also know  that such corrections change the value and energy dependence of the saturation scale
\cite{MUTR,DG}. Therefore, it is necessary    to include the NLO correction to obtain reliable estimates of the non-linear effects.

It is  not sufficient to solve the Balitsky-Kovchegov (BK)   equation because together with the NLO corrections to the Balitsky-Kovchegov equation, we need to include in the calculating procedure the corrections referred to as Pomeron loops\cite{MUSH,MSHW,LELU,KOLU,HIMST,AKLP,MPSI,LMP}.  In spite of the fact, that we have learnt a lot about these corrections, no closed equation  exists  and the whole procedure turns out to be so complicated, that there is no hope of obtaining   reliable estimates in the foreseeable  future.

In such a situation the only thing that we can do,  is to try to study the influence of different terms in NLO corrections to the Balitsky-Kovchegov equation,  with the goal of finding  the  essential terms. After that  we can  try to estimate the influence of the Pomeron loops and  find out what is more important: loops or NLO corrections. We are presently at the first stage of such a study.

The clear example of such approach is the effect of running QCD  coupling constant. The running QCD coupling constant is believed to be  one of the most characteristic features of QCD, and it is important to find out how this phenomenon  affects the value of the scattering amplitude. Fortunately, the form of the BK equation is known \cite{RAS} and the numerical solution\cite{RASNUM}  shows
that the solution with running QCD coupling constant  is quite different from the one with fixed coupling.

The second example is the solution to more general JIMWLK equation which includes all $1/N_c$ correction to Balitsky-Kovchegov equation (see Ref.\cite{NCCOR}). It turns out that the difference, between  the solution  of the JIMWLK equation and the  BK one, is extremely small.

In this paper we wish to investigate the influence of  the conservation of energy on the solution to the BK equation. As has been discussed the important observable for the interpretation of the nucleus-nucleus collision is the energy loss\cite{ELOSS}. In LO the non linear equation is written neglecting the energy loss. Therefore, we need to generalize the BK equation in such way that it would respect  energy conservation. Such a modified equation will allows us to discuss the value and energy and transverse momentum dependence of the nucleus modification factor(NMF). Since for  single inclusive production the $k_t$ factorization has been proven (see Ref.\cite{KTINC})  the value of the NMF, that is originated from short distances, is determined by the BK equation. The generalized  BK equation has been suggested in Ref.\cite{MBK} and for the completeness of presentation we will discuss this equation in the next section.
The main section of the paper is the third one in which we present the solution to this equation closely
following  the method proposed in Ref.\cite{LT}.  We find no
 considerable effects due to energy conservation,  and show that the dipole scattering amplitude in the region $r > 1/Q_s$   leads to the inclusive gluon production that is   20 - 40\%  smaller than that is predicted  from the   LO BK  equation.

The second problem that  we discuss in this paper is the impact parameter dependence of the solution both to Balitsky-Kovchegov equation and to the modified version of this equation. It is well known that $b$ dependence is one of the most  difficult and challenging problems  in  QCD\cite{KOWI}.  The massless gluon leads  to power-like decrease of the scattering amplitude for large values
of $b$ which results in the power-like increase of the interaction radius\cite{KOWI}. Such behavior of the interaction radius indicates that the main contribution to the solution of the non-linear equation stems from large values of $b$ making the entire approach, based on perturbative QCD,  inconsistent. The attempts ( only two \cite{BDEP} as far as we know)  to solve BK equation numerically taking into account $b$ dependence, confirm these  pessimistic expectations. The simplified  version of BK equation allows one  to obtain the analytical solution including $b$-dependence. It turns out that  the non-perturbative $b$ dependence can be absorbed in the saturation scale.  This observation gives us a guide for the solution of the impact parameter problem in high density QCD, but how  to include this observation into general BK equation still remains  unsolved. In section 4 we discuss the inclusive production for the solution with impact parameter dependence.

\section{The modified non-linear equation}
\setcounter{equation}{0}

The non-linear (BK) equation has the following form for the dipole scattering amplitude $N\Lb r,Y;b \Rb$
\bea \label{BK}
\frac{\partial N\Lb r,Y;b \Rb }{\partial\,Y}\,\,
&=&\,\,\frac{\bas}{\pi}\,\,\int\,d^2 r'\,K\Lb  r,r'\Rb\,
\Lb \,2  N\Lb r',Y;\vec{b} \, -\, \frac{1}{2}\,(\vec{r} - \vec{r}\:')\Rb\,\,- \,\, N\Lb r,Y;\vec{b}\Rb\,\,\right.\nonumber\\
&-& \left.  \,\,
 N\Lb r',Y;\vec{b} - \frac{1}{2}\,(\vec{r} - \vec{r}\:') \Rb\, N\Lb
 \vec{r} -
\vec{r}\:',Y;\vec{b} - \frac{1}{2} \vec{r}\:'\Rb  \,
\Rb
\eea
where $Y = \ln(1/x)$,$\bas =\frac{C_F\,\as}{\pi}$, $x$
 denotes the Bjorken variable for the dipole-target scattering, and $b$
is the impact parameter for the reaction. Kernel $K\Lb r,r'\Rb $ in the LO is equal to
\beq \label{KLO}
K\Lb r,r'\Rb \,\,\,=\,\,\,
\frac{r^2}{(\vec{r}\,-\,\vec{r}\:')^2\,r'^2}
\eeq

\subsection{The linear equation in NLO}


The linear part of \eq{BK} can be easily written in the double Mellin transform
\beq \label{DMELL}
N \Lb \xi \,=\,\ln(r^2/R^2),Y = \ln(1/x); b \Rb \,=\,\,\int \,\frac{d \omega}{2 \pi\,i}\,\frac{d \gamma}{2
\pi\,i}\,N(\omega,\gamma; b)\,e^{ \omega\,Y\,\,+\,\,\gamma\,\xi}
\eeq
It has the form
\beq \label{BFKLOM}
\omega\,N(\omega,\gamma; b)\,=\,\Lb \bas\,\chi_{LO}(\gamma)\,+\,\bas^2\,\chi_{NLO}(\gamma) \Rb\,N(\omega,\gamma; b) \,\,\,\mbox{or}\,\,\,\omega\,\,=\,\,\omega(\gamma)\,\,=\,\,\bas\,\chi_{LO}(\gamma)\,\,+\,\,\bas^2\,\chi_{NLO}(\gamma)
.
\eeq
 For $\chi_{LO}(\gamma)$ we have the well known expression
\cite{BFKL}:

\beq \label{CHILO}
\chi_{LO}(\gamma)\,\,=\,\,2\,\psi(1) \,\,-\,\,\psi(\gamma)\,\,-\,\,\psi(1 - \gamma)\,\,=\,\,\frac{1}{\gamma}
 + \frac{1}{ 1 - \gamma}
\,\,+\,\,\chi^{HT}_{LO}(\gamma),
\eeq

with $\psi = d \ln \Gamma (\gamma)/d \gamma$ and $\Gamma$
 is  the Euler Gamma function. We will discuss the form of
$\chi_{NLO}(\gamma)$
 later. $\chi^{HT}_{LO}(\gamma) = 2\,\psi(1) \,\,-\,\,\psi(1 +
 \gamma)\,\,-\,\,\psi(2 - \gamma)$ denotes  the contribution of the higher
 twist which we will discuss below.

The NLO BFKL kernel is known \cite{NLLG,FG}, following Ref.  \cite{MBK} we use the simplified form of the NLO kernel:
\begin{itemize}
\item\quad In Ref.\cite{DG}  the following form of the NLO BFKL kernel is proposed
\beq \label{NLOKDG}
\bas\,\chi_{NLO}(\gamma)\,\,\,=\,\,\,\Lb
\frac{1 + \omega\,A_1(\omega)}{ \gamma} - \frac{1}{\gamma} + \frac{1 +
\omega\,A_1(\omega)}{1 -  \gamma + \omega} - \frac{1}{ 1 - \gamma}\,\Rb \,\,-\,\,\omega \,
\chi^{HT}_{LO}(\gamma)\,,
\eeq
where,
\beq \label{AOM}
A(\omega) = - 11/12 + O(\omega) + n_F \Lb
\frac{\bas}{4\,N^2_c\,\gamma}\,P_{Gq}(\omega)\,P_{qG}(\omega) - 1/3 \Rb\,,
\eeq
with $P(\omega)$ being the DGLAP kernel.

\item\quad
The singularities in \eq{NLOKDG} describe the different branches of evolution
 corresponding to
 the sizes of the interacting dipole (or the  transverse momenta of
 partons). The pole at $\gamma=0$ corresponds to the normal twist-2 DGLAP
 contribution, with the ordering in the transverse parton momenta $Q >
\dots
 k_{i,t} > k_{t,i+1} > \dots > Q_0$, where $Q_0$ is the typical
virtuality of
 the target $Q_0 \approx \,1/R$. The pole at $\gamma = 1$ in \eq{NLOKDG}
 corresponds to inverse $k_t$ ordering ($ k_{i,t} < k_{t,i+1} < \dots
 \,Q_0$).  The other poles, at $\gamma = -1, -2, \dots (\gamma = 2,3,
 \dots)$, are the higher twists contributions due to the gluon reggeization.

\item\quad
As can be seen in \eq{NLOKDG}, the main changes in the NLO kernel   to the ordinary DGLAP evolution are introduced,
so as to account for the DGLAP anomalous
dimension,  and to the part of the kernel that describes
 the
 inverse evolution. This branch of evolution is moderated by the
 non-linear effect, rather than by the NLO corrections.  Indeed, it was
shown
 in Ref. \cite{LT} that the term $1/(1 - \gamma)$ leads to  exponentially
 small corrections in the saturation region. Consequently  we can neglect changes in the inverse evolution, and
 keep the BFKL kernel without a shift from $\gamma $ to $\gamma - \omega$.
\item\quad  We use
the observation of Ref. \cite{EKL}, according to which \eq{AOM} for
 $A_1(\omega)$ can be  approximated to an   accuracy ($> 95\%$),
by
 $A_1(\omega) = 1$.

\end{itemize}
Hence, we  can  rewrite the NLO BFKL kernel in a very
simple form:
\beq \label{NLOOUR}
\chi_{NLO}(\gamma) \,\,\,=\,\,\, - \omega\,\chi_{LO}(\gamma)\,.
\eeq

Using \eq{NLOOUR} we obtain the full kernel for the linear equation in the
 form:
\beq \label{KERF} \omega(\gamma) \,\,=\,\,\bas \,\chi_{LO}(\gamma)\,\Lb
 1 \,\,-\,\,\omega \Rb
\eeq

This kernel imposes  energy conservation (see Ref. \cite{EKL}), and
 describes the NLO BFKL kernel. It does not include  the contribution
coming from
 inverse ordering, which should be suppressed in the solution to the
 non-linear equation.

 Note  that  substituting $\chi_{LO} = \bas/\gamma$  yields
\beq \label{ADEKL}
 \gamma(\omega)\,\,=\,\,\bas\,\left(\,\frac{1}{\omega}\,\,-\,\,1\,\right)\,,
 \eeq
 which approximates the DGLAP anomalous dimension in  leading order so well, that the difference between this
anomalous dimension and \eq{ADEKL} turns out to be less than
5\% \cite{EKL}.

 \eq{ADEKL} corresponds to \eq{NLOKDG}  where  $A_1(\omega) = 1$, and  the high twist terms are neglected.

\subsection{The non linear equation in NLO}


The main idea of Ref.\cite{MBK} is to use  \eq{KERF} as the kernel for the non-linear equation.

First one can see that  \eq{KERF} can be rewritten in the coordinate space as

$$
\frac{\partial N\Lb r,Y;b \Rb}{\partial\,Y}\,\,\,
=\,\,\frac{C_F\,\as}{\pi^2}\,\,\int\,d^2 r'\,K_{LO}\Lb r,r'\Rb\,\Big\{
1\,\,\,-\,\,\frac{\partial}{\partial Y}\Big \}
$$
\beq \label{MODBK}
\Lb \,2  N\Lb r',Y;\vec{b} \, -\
\frac{1}{2}\,(\vec{r} - \vec{r}\:')\Rb\,\,- \,\, N\Lb r,Y;\vec{b}\Rb\,\,
\Rb.
\eeq

As  is shown in Refs. \cite{MUCD,MPSI,LLA} the non linear evolution equation has a very simple probabilistic interpretation and can be written as the equation for the Markov chain or, in other words, can be viewed as the process of the birth (non linear term) and death (linear term) of the colorless dipoles.  Assuming that this interpretation is correct in NLO order, we obtain the following equation\cite{MBK}:
\bea
&&\frac{\partial N\Lb r,Y;b \Rb}{\partial\,Y}\,\,\,
=\,\,\frac{C_F\,\as}{\pi^2}\,\,\int\,d^2 r'\,K_{LO}\Lb r,r'\Rb\,\Big\{
1\,\,\,-\,\,\frac{\partial}{\partial Y}\Big \}\label{MODBKF}\\
&&
\Lb \,2  N\Lb r',Y;\vec{b} \, -\
\frac{1}{2}\,(\vec{r} - \vec{r}\:')\Rb\,\,- \,\, N\Lb r,Y;\vec{b}\Rb\,\,-\,\,
N\Lb r',Y;\vec{b} - \frac{1}{2}\,(\vec{r} - \vec{r}\:') \Rb\, N\Lb
 \vec{r} -
\vec{r}\:',Y;\vec{b} - \frac{1}{2} \vec{r}\:'\Rb
\Rb \nn
\eea
 The advantage of this equation is that energy is conserved by the non-linear
 term, as well as by the linear one. The linear term has been discussed above. We need only to rewrite the non-linear term in
Mellin transform ( see \eq{DMELL}). namely,
\bea \label{N2OM}
&&N\Lb r',Y;\vec{b} - \frac{1}{2}\,(\vec{r} - \vec{r}\:') \Rb\, N\Lb
 \vec{r} -
\vec{r}\:',Y;\vec{b} - \frac{1}{2} \vec{r}\:'\Rb\,\,=\\
&&\,~~~~~~~~~~~~~~~~~~~~~~~~\,\,\int\,\frac{d \omega\,d \omega'}{( 2 \pi i)^2}\,e^{( \omega - \omega')Y +
\omega' Y}\,{\cal N}\Lb r',\omega - \omega';\vec{b} - \frac{1}{2}\,(\vec{r} - \vec{r}\:') \Rb\, {\cal N}\Lb
 \vec{r} -
\vec{r}\:',\omega';\vec{b} - \frac{1}{2} \vec{r}\:'\Rb \nn
\eea
to understand why this term conserves energy, as the derivative with respect to  $Y$ leads to extra factor $\omega$ in the
 non-linear term.
\section{Solution to the modified B-K equation}

\subsection{Simplified kernel}

\subsubsection{General approach}

As was suggested in Ref.\cite{LT} we replace the full BFKL kernel in LO by the following simplified kernel
\bea \label{GALO}
\omega_{LO}\Lb \gamma\Rb\,\,=\,\,\bas \left\{\begin{array}{l}\,\,\,\frac{1}{\gamma}\,\,\,\,\,\mbox{for}\,\,\,z\,=\,\ln\Lb r^2 Q^2_s\Rb\,\leq \,0\,;\\ \\
\,\,\,\frac{1}{1 \,-\,\gamma}\,\,\,\,\,\mbox{for}\,\,\,z\,=\,\ln\Lb r^2 Q^2_s\,\Rb\,>\,0\,; \end{array}
\right.
\eea
This kernel leads to the leading twist contribution and gives the natural generalization of the DGLAP equation  that includes two different kind of logs: for  $z <0$ it sums $\Lb\bas\ln\Lb  r^2 \Lambda_{QCD}\Rb\Rb^n$ while for $z>0$
the new type of  appears, namely, $\Lb\bas\ln\Lb  r^2  Q^2_s\Rb\Rb^n$.

For the NLO kernel we assume that for $z\,=\,\ln\Lb r^2 Q^2_s\Rb\,\leq \,0\,$ we have \eq{ADEKL} which can be rewritten as
\beq \label{NLO1}
\omega_{NLO}\Lb \gamma\Rb\,\,=\,\,\bas \frac{1}{\gamma + \bas}
\eeq
This form of  the kernel shows that in the NLO,  we sum, in addition to  $\Big(\bas \ln\Lb 1/\Lb r^2\,\Lambda_{QCD}\Rb\Rb\Big)^n$
terms, also non-logarithmic corrections. Recall that log contribution stems from the singularity $ (1/\gamma)$ in the kernel.

For $z\,=\,\ln\Lb r^2 Q^2_s\Rb\,\geq \,0\,$ we parameterize the NLO kernel  as the solution to the following equation:
\beq \label{NLO2}
\omega_{NLO}\Lb \gamma\Rb\,\,=\,\,\bas\,\Lb 1\,-\,\omega_{NLO}\Lb \gamma\Rb\Rb\,\Big\{ \frac{1}{1\,-\,\gamma}\,\,+\,\,\as\,\kappa\Big\}
\eeq
where we determine the  constant $\kappa$ from the condition
\beq \label{NLO3}
\omega_{NLO}\Lb \gamma \,\,\xrightarrow{\gamma \geq \gamma_{cr}} \,\,\gamma_{cr}\Rb\,\,\,=\,\,\,
\omega_{NLO}\Lb \gamma \,\,\xrightarrow{\gamma \leq \gamma_{cr}} \,\,\gamma_{cr}\Rb
\eeq

Therefore, to find $\kappa$ it is necessary  to discuss the critical value of the anomalous dimension.
\subsubsection{Critical anomalous dimension and saturation momentum}
Using \eq{GALO}
one can calculate the critical anomalous dimension. It is well known that for this calculation one only  needs to know  the kernel for $z <0$( see Refs.\cite{GLR,BALE,MUTR,MUPE}). The equation for the critical anomalous dimension ($\gamma_{cr}$)   has the  following form
\beq \label{GACREQ}
- \frac{\partial \omega(\gamma_{cr})}{ \partial \gamma_{cr}}\,\,=\,\,\frac{\omega(\gamma_{cr})}{ 1 - \gamma_{cr}}
\eeq
Inserting \eq{GALO} in \eq{GACREQ} one obtains
\beq \label{GACR}
\gamma_{cr}\,\,=\,\,\frac{1}{2}\,\Lb 1 \,-\,\bas\Rb
\eeq
he equation for the saturation momentum has the form\cite{GLR,BALE,MUTR,MUPE}:
\beq \label{QS}
\ln\Lb Q^2_s/Q^2_0\Rb \,\,=\,\,\frac{4 \bas}{\Lb 1 + \bas\Rb^2}\,Y
\eeq
where $Y = \ln(1/x)$, and $Q_0$ is the scale  associated with the initial condition at low energies.

In the vicinity of the saturation scale $ z \to 1$ the behavior of the dipole amplitude has the form \cite{MUTR,IIM}
\beq \label{VICQS}
N\Lb Y; r\Rb\,\,\propto \Lb r^2 Q^2_s\Rb^{ 1 - \gamma_{cr}}\,\,=\,\,e^{\Lb 1 - \gamma_{cr}\Rb z}
\eeq

\begin{boldmath}
 \subsubsection{The NLO kernel for $\gamma \geq \gamma_{cr}$}
\end{boldmath}
Using \eq{GACR} we can calculate

\beq \label{NLO4}
\omega_{NLO}\Lb \gamma \,\,\xrightarrow{\gamma \leq \gamma_{cr}} \,\,\gamma_{cr}\Rb\,\,\,=\,\,\frac{2 \,\bas}{ 1 \,+\,\bas}
\eeq
and \eq{NLO3} leads  to the  following value of $\kappa$:
\beq \label{NLO5}
\kappa\,\,=\,\,\frac{4 \,\bas}{ 1  \,-\,\bas^2}
\eeq
Finally, the NLO kernel has the form:
\bea \label{NLO6}
\omega_{NLO}\Lb \gamma\Rb\,\,=\,\,\bas \left\{\begin{array}{l}\,\,\,\frac{1}{\gamma\,+\,\bas}\,\,\,\,\,\mbox{for}\,\,\,z\,=\,\ln\Lb r^2 Q^2_s\Rb\,\leq \,0\,;\\ \\
\,\,\,\frac{1\,-\,\gamma \,+\,\bas\,\kappa}{\Lb 1 \,-\,\gamma\Rb\,\Lb 1 \,+\,\bas\,\kappa\Rb \,+\,\bas}\,\,\,\,\,\mbox{for}\,\,\,z\,=\,\ln\Lb r^2 Q^2_s\,\Rb\,>\,0\,; \end{array}
\right.
\eea

It is worthwhile mentioning that the kernel of \eq{NLO6} describes the LO anomalous dimension of the DGLAP equations for $z \geq 0$,  and  it takes into account  energy conservation.  Therefore, we consider it as a model that can lead to a reliable predictions.

\subsection{The simplified equation in the coordinate space}

\begin{boldmath}
\subsubsection{z $<$  0}
\end{boldmath}

 In this kinematic region we can simplify  $K_{LO}\Lb r,r'\Rb$ in \eq{MODBKF}  in the following way\cite{LT}, since $r'
 \gg r$ and $|\vec{r} - \vec{r}'| > r$
 \beq \label{SEQCS1}
 \int d^2 r' \,K_{LO}\Lb r, r'\Rb\,\,\rightarrow\,\pi\, r^2\,\int^{\frac{1}{\Lambda^2_{QCD}}}_{r^2} \frac{ d r'^2}{r'^4}
\eeq
Introducing $n \Lb r,Y;b\Rb= N\Lb r,Y;b\Rb/r^2$ we obtain
 \beq \label{SEQCS2}
\frac{\partial^2 n\Lb r,Y;b\Rb}{\partial Y\,\partial \ln\Lb 1/(r^2 \Lambda^2_{QCD})\Rb}\,\,\,=\,\,\bas\Big\{ 1 \,-\,\frac{\partial}{\partial Y}\Big\}\,\Big( 2 n\Lb r,Y;b\Rb\,\,- \,\,r^2\Lambda^2_{QCD} n^2\Lb r,Y;b\Rb\Big)
\eeq

\begin{boldmath}
\subsubsection{z $>$  0}
\end{boldmath}

The main contribution in this kinematic region originates from the decay of the large size dipole into one small size dipole  and one large size dipole.  However, the size of the small dipole is still larger than $1/Q_s$. This observation can be translated in the following form of the kernel
\beq \label{SEQCS3}
 \int d^2 r' \,K_{LO}\Lb r, r'\Rb\,\,\rightarrow\,\pi\, \int^{r^2}_{1/Q^2_s(Y,b)} \frac{ d r'^2}{r'^2}\,\,+\,\,
\pi\, \int^{r^2}_{1/Q^2_s(Y,b)} \frac{ d |\vec{r} - \vec{r}'|^2}{|\vec{r} - \vec{r}'|^2}
\eeq
One can see that this kernel leads to the $\ln\Lb r^2Q^2_s\Rb$-contribution. Introducing a new function $\tilde{N}\Lb r,Y;b\Rb\,\,=\,\,\int^{ r^2} d r^2\,N\Lb r,Y;b\Rb/r^2$ one obtain the following equation
\beq \label{SEQCS4}
\frac{\partial^2 \tilde{N}\Lb r,Y;b\Rb}{ \partial Y\,\partial \ln r^2}\,\,=\,\,2 \bas \Big\{ 1\,-\,\frac{\partial}{\partial Y}\Big\}\,\Lb \Lb 1 \,+\,\bas\,\kappa\,\frac{\partial}{\partial \,\ln r^2}\,\,-\,\frac{\partial \tilde{N}\Lb r,Y;b\Rb}{\partial  \ln r^2}\Rb \, \tilde{N}\Lb r,Y;b\Rb\Rb
\eeq
This equation does not depend on the value of the saturation momentum. Because of this we can assume that the solution of this equation depends  on one variable $z = \ln\Lb r^2 Q^2_s\Rb$. The result that the solution in the saturation region depends on only one variable, has been proven (see Ref.\cite{BALE}).
The value of the saturation momentum depends on two variables: on rapidity  $Y$ and impact parameter $b$. Since the equation does not depend explicitly on $Q_s$ and $b$, the only information that we have on these variables stems from the matching with the equation at $z < 0$ on the critical line.
Transforming  from variable $Y$ and $r$ in \eq{SEQCS4} to variable $z = \ln\Lb r^2 Q^2_s\Rb$ we have
\beq  \label{SEQCS5}
\frac{d^2  \tilde{N}\Lb z\Rb}{d z^2}\,\,=\,\,\frac{(1 + \bas)^2}{4}\,
\Lb 1\,+\,\frac{4\,\bas}{1\,-\,\bas^2}\,\frac{d}{d z}\, -\, \frac{d  \tilde{N}\Lb z\Rb}{ d z} \Rb  \tilde{N}\Lb z\Rb\,\,-\,\,\bas \frac{d}{d z} \Big\{\Lb 1 - \frac{d  \tilde{N}\Lb z\Rb}{ d z} \Rb \tilde{N}\Lb z\Rb\Big\}
\eeq

It worthwhile mentioning that the BK equation in this kinematic region has the form
\beq  \label{SEQCS6}
\frac{d^2  \tilde{N}\Lb z\Rb}{d z^2}\,\,=\,\,\frac{1}{4}\,
\Lb 1 - \frac{d  \tilde{N}\Lb z\Rb}{ d z} \Rb  \tilde{N}\Lb z\Rb\,\,
\eeq


\subsection{Solution to BK equation with the simplified kernel}


For the sake of completeness we start by recalling the solution to BK equation  in the form of \eq{SEQCS6}, obtained in Ref. \cite{LT}.
Introducing
\beq \label{S1}
\tilde{N}\Lb z\Rb\,\,=\,\,\int^z\,d z' \,\Big( 1\,\,-\,\,e^{- \phi\Lb z' \Rb}\Big)
\eeq
 \eq{SEQCS6} can be rewritten in the form
\beq \label{S2}
\frac{d  \phi\Lb z\Rb}{ d z}\,\,=\,\,\frac{1}{4}\,\int^z\,d z' \,\Big( 1\,\,-\,\,e^{- \phi\Lb z' \Rb}\Big)
\eeq
Considering $d  \phi\Lb z' \Rb/ d z = D\Lb \phi\Rb$  we can rewrite \eq{S2} as
\beq \label{S3}
 D\Lb \phi\Rb\,\,=\,\,\frac{1}{4}\,\int^{\phi}_{\phi_0}\,\frac{d \phi'}{D\Lb \phi'\Rb}\,\Big( 1\,\,-\,\,e^{- \phi'}\Big)
\eeq
Differentiating \eq{S3} with respect to $\phi$ we have
\beq \label{S4}
\frac{d D\Lb \phi\Rb}{ d \phi} \,\,=\,\frac{1}{4}\,\frac{1}{D\Lb \phi\Rb}\,\Big( 1\,\,-\,\,e^{- \phi}\Big)
\eeq
which leads to
\beq \label{S5}
 D^2\Lb \phi\Rb\,\,=\,\,\frac{1}{2}\,\Big(\phi \,+ \,e^{ - \phi} - 1\Big)
\eeq
and finally,
\beq \label{S6}
z\,\,=\,\,\sqrt{2}\,\int^\phi_{\phi_0} \frac{d \phi'}{\sqrt{\phi \,+ \,e^{ - \phi} - 1}}
\eeq

The boundary conditions at $z = 0$ have the form\cite{GLR,LT,IIM}
\bea
\phi\Lb z\Rb \,\xrightarrow{z \to  0^-}\,\,&\phi_0 &\,\,\,\xleftarrow{z \to  0^+}\,\,\,\phi\Lb z\Rb  ;\label{BC01}\\
\frac{ d \ln\Lb \phi\Lb z\Rb\Rb}{d z}\,\,\xrightarrow{z \to  0^-}\,\,\,&\frac{1}{2}&\,\,\,\xleftarrow{z \to  0^+}\,\,\frac{d ln\Lb \phi\Lb z\Rb\Rb}{d z}\,;\label{BC02}
\eea
\eq{BC02} follows from \eq{VICQS}  for the kernel given by \eq{GALO}

\subsection{Solution to the modified BK equation}

Using \eq{S1} we can reduce \eq{SEQCS5} to the form
\beq \label{S7}
\phi'_z\,e^{- \phi(z)}\,
\,=\,\,\frac{( 1 + \bas)^2}{4}\,e^{- \phi(z)}\,\tilde{N}\Lb z \Rb\,\,+\,\,\frac{\bas\,\Lb 1  + \bas\Rb}{1\,-\,\bas}\,\tilde{N}'_z\,\,-\,\,\bas\Big( e^{- \phi(z)}\,\tilde{N}\Lb z \Rb\Big)'_z
\eeq
or
\beq \label{S8}
\phi'_z \,\Lb 1\,-\,\bas\, \tilde{N}\Rb\,\,=\,\,\frac{( 1 + \bas)^2}{4}\,\tilde{N}\Lb z \Rb\,+\,\bas\,\Lb 1\,-\,\,e^{ - \phi\Lb z\Rb}\Rb^2
\eeq
Considering $\phi$ being function on $\tilde{N}$ and neglecting the terms that are proportional to $\bas^2$, we can rewrite \eq{S8} as
\beq \label{S10}
\phi'_{\tilde{N}}\,\,\,=\,\,\frac{1}{ 1 \,-\,\bas\,\tilde{N}}\,\Big(\,\frac{(1 + \bas)^2}{4}\,\frac{\tilde{N}}{1\,-\,\exp\Big\{ - \phi\Lb \tilde{N}\Rb\Big\}}\,\,\,\,+\,\,\,\,\bas\,\Big( 1\,-\,\exp\Big\{ - \phi\Lb \tilde{N}\Rb\Big\} \Big) \Big)
\eeq
The BK equation for $ \phi\Lb\tilde{N}\Rb$ has the form
\beq \label{S101}
\phi'_{\tilde{N}}\,\,\,=\,\,\frac{1}{4}\,\frac{\tilde{N}}{1\,-\,\exp\Big\{ - \phi\Lb \tilde{N}\Rb\Big\}}
\eeq

In comparison with the BK equation  \eq{S10}  (see \eq{S101}) this contains a new feature: $ \phi'_{\tilde{N}} \to \infty$ at $\tilde{N} \to 1/\bas$. In the vicinity of $\tilde{N} = 1/\bas$ we can simplify \eq{S10}, namely,
\beq \label{S11}
\phi'_{\tilde{N}}\,\,\,=\,\,\frac{1}{ 1\,-\,\bas \tilde{N}}\,\,\Big(\,\frac{(1 + \bas)^2}{4}\,\frac{\tilde{N}}{1\,-\,\exp\Big\{ - \phi\Lb \tilde{N}\Rb\Big\}}\Big)
\eeq
since the first term in the bracket in \eq{S10} is of the order of $ \geq \,1/\bas$.

The solution to \eq{S11} has the form
\beq \label{S12}
\phi \,\,+\,\,e^{ - \phi}\,\, -\,\, 1 \,\,=\,\,\frac{(1 + \bas)^2}{4}\,\frac{1}{\bas^2}\Big( \ln \Big(1\,-\,\bas \tilde{N}\Big)\,\,+\,\,\bas\Big)
\eeq
 From   \eq{S12} we can see that $\phi$ is large in the region where $\bas \tilde{N} \,\ll\,\,1$. Having this in mind we can replace \eq{S10} by the following equation
\beq \label{S13}
\phi'_{\tilde{N}}\,\,\,=\,\,\frac{1}{ 1\,-\,\bas \tilde{N}}\,\,\Big(\,\frac{(1 + \bas)^2}{4}\,\tilde{N}\,\,\,\,+\,\,\,\,\bas \Big)
\eeq
which gives
\beq \label{S14}
\phi \Lb \tilde{N}\Rb\,\,=\, \,\frac{1}{\bas^2}\,\Big( - \bas \frac{(1 + \bas)^2}{4} \,\tilde{N}\,\,\,-\,\,\,\Lb \bas^2 \,+\,\frac{(1 + \bas)^2}{4}\Rb \,\ln\Lb 1\,-\,\bas \tilde{N}\Rb
\eeq

\FIGURE[h]{\begin{minipage}{60mm}{
\centerline{\epsfig{file=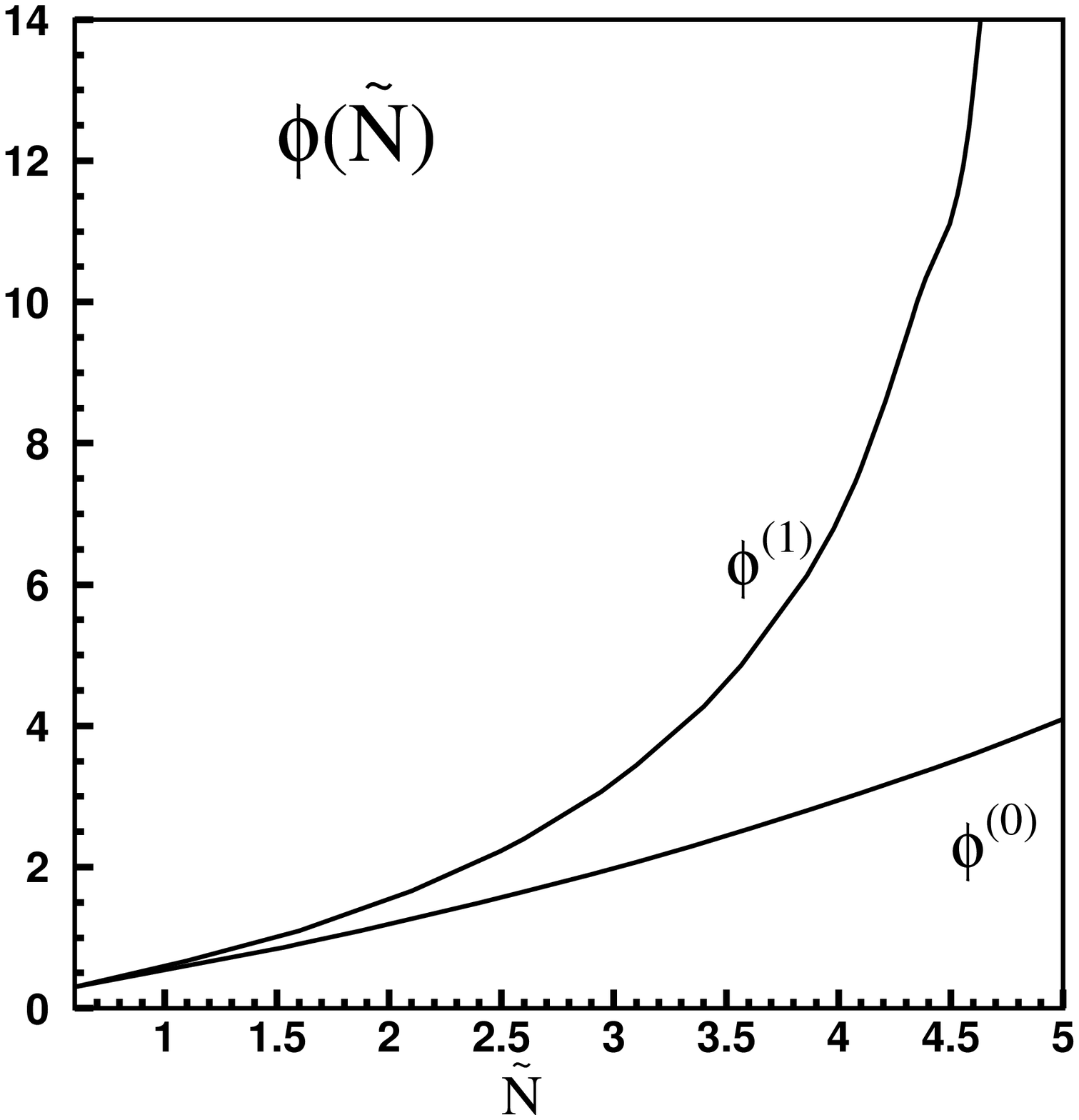,width=50mm}}
\caption{Dependence of $\phi\Lb \tilde{N}\Rb$ on $\tilde{N}$ for solution to  \eq{S10} ($\phi^{(1)}$ and to \eq{S101} ($\phi^{(0)}$.  $\bas = 0.2$,$\phi_0=0.3$, $ \tilde{N}_0 = 0.6$}}
\end{minipage}
\label{phn}}

In \fig{phn} we illustrate  the behavior of $\phi \Lb \tilde{N}\Rb$ for the  solution to \eq{S10} and \eq{S101}.  The difference between
modified equation and BK one is clearly seen,  and it is well reproduced by the simplified solutions of \eq{S12} and \eq{S14}.
However, for small values of $\tilde{N}$ we have to solve equation numerically to determine  the accuracy of the estimates.
Having $\phi\Lb \tilde{N}\Rb$ we can find $\tilde{N}$ as function of $z$. Indeed,
\beq \label{S15}
\frac{d \tilde{N}\Lb z \Rb}{d z} \,\,=\,\,1\,-\,\exp\Lb - \phi\Lb \tilde{N}\Rb\Rb
\eeq
and the following equation gives $\tilde{N}$ as function of $z$
\beq \label{S16}
\int^{\tilde{N}}_{\tilde{N}_0}\,\frac{d \tilde{N}'}{1\,-\,\exp\Lb - \phi\Lb \tilde{N}'\Rb\Rb}\,\,\,=\,\,z
\eeq
The value of $\tilde{N}\Lb z=0\Rb = \tilde{N}_0$ can estimated from \eq{VICQS} . Indeed, in Ref.\cite{LT,IIM} was shown that for  $z < 0$ but in the vicinity of $z = 0$ $\tilde{N} = \int^z_{-\infty} d z' \Big( 1 - \exp\Lb - \phi_0\,\exp\Lb (1 - \gamma_{cr}) z\Rb\Rb\Big)$ . Taking this integral we obtain the value for $\tilde{N}_0$.

\FIGURE[h]{\begin{tabular}{ c c }
\epsfig{file=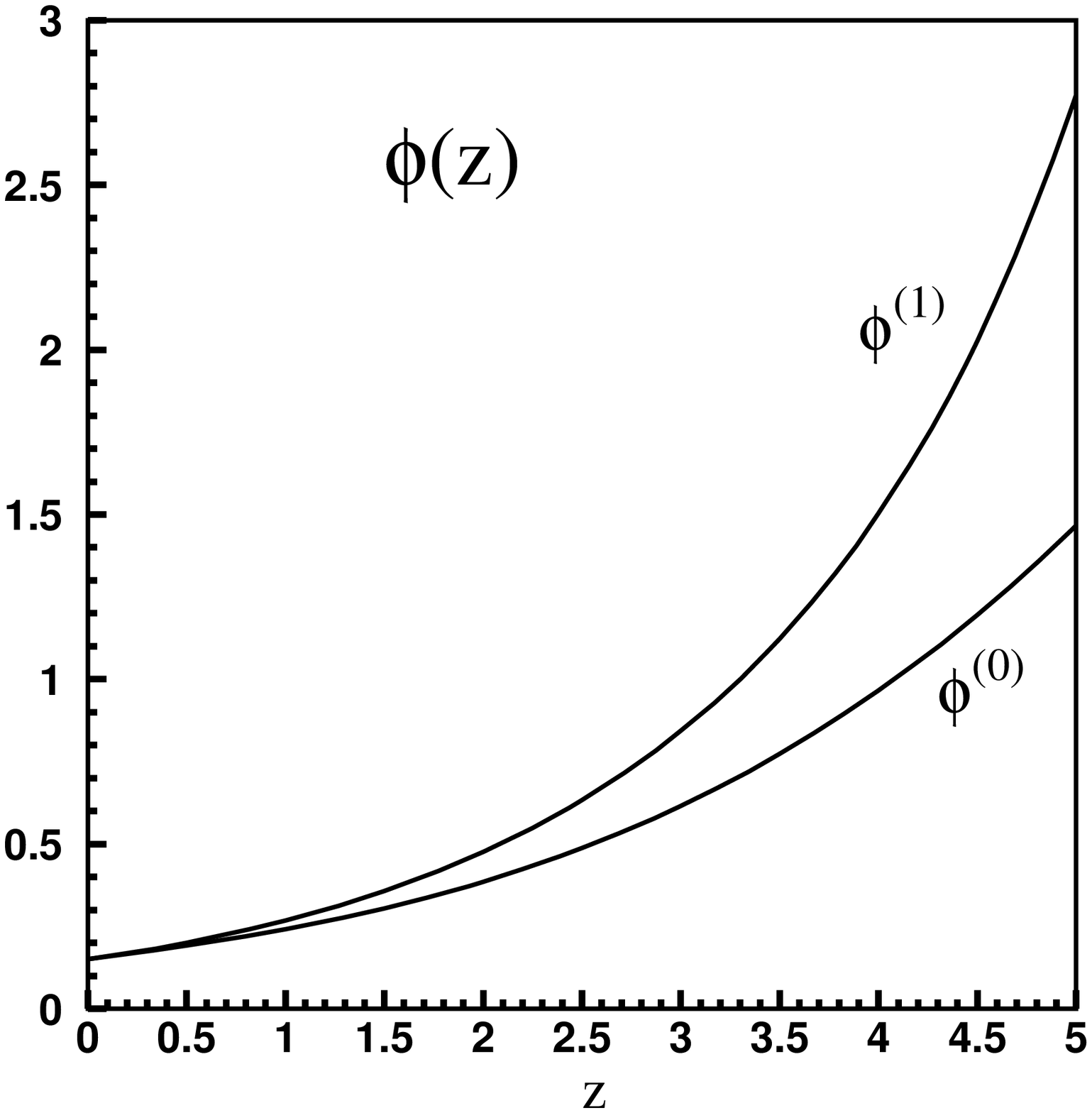,width=90mm} &\epsfig{file=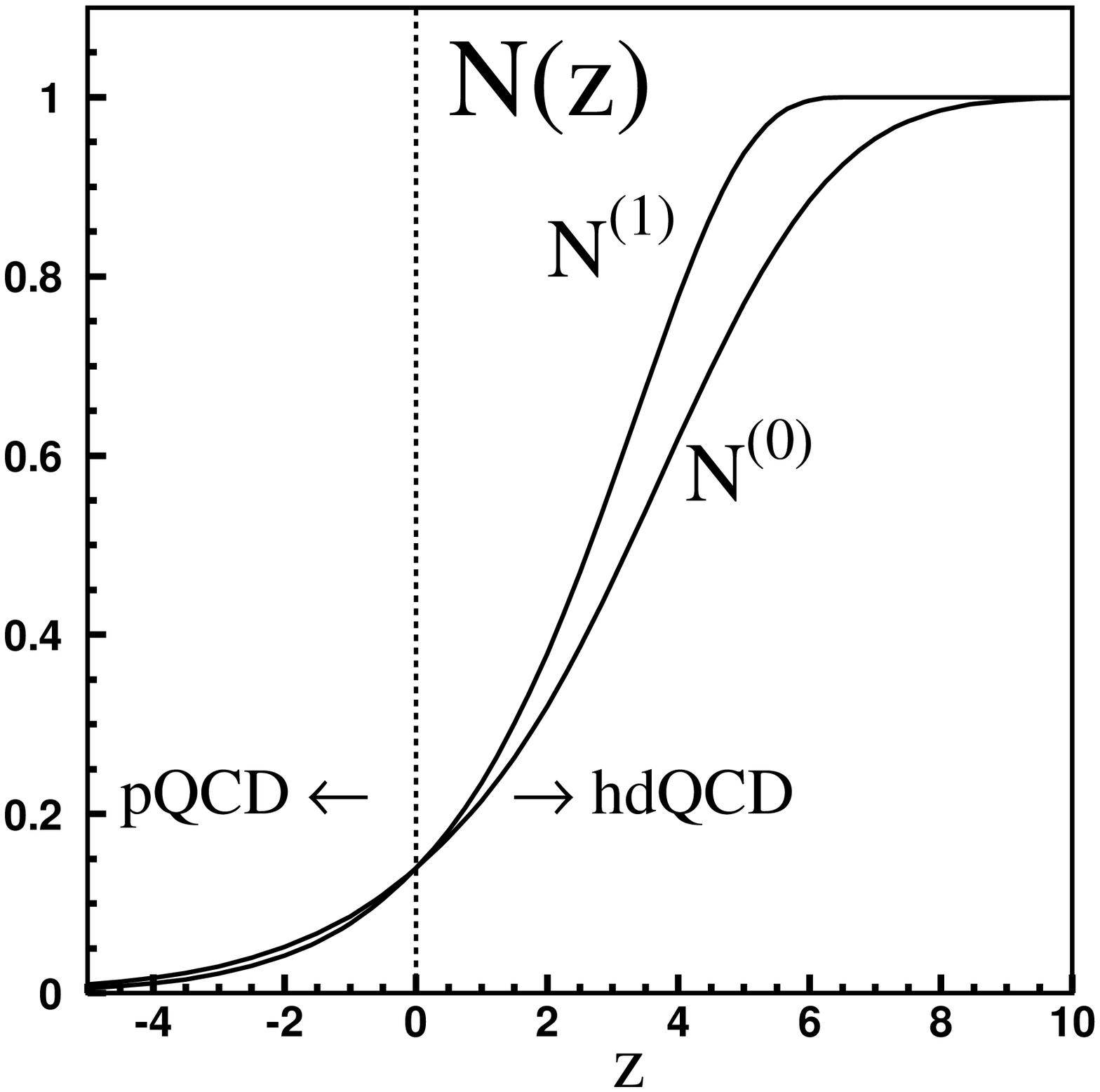,width=90mm}\\
\fig{nz} - a & \fig{nz}-b
\caption{Dependence of  function $\phi(z)$ (\fig{nz}-a) and the
scattering amplitude $N(z)$( \fig{nz}-b)  versus $z$  for solution to  \eq{S10} ($N^{(1)}$) and to \eq{S101} ($N^{(0)}$).  $\bas = 0.2$,$\phi_0=0.3$, $\tilde{N}_0 = 0.6$}
\label{nz}
\end{tabular}}
In \fig{nz} the solutions for modified  equation and for the BK equation  are plotted. The comparison between them is  illustrated in \fig{nzr}  in which the ratio $N^{(1)}/N^{(0)}$ is shown, where $N^{(1)}$ and $N^{(0)}$ are solutions to \eq{S10} and \eq{S101}, respectively. One can see that the effect could be as large as 20\% (see \fig{nzr})

\subsection{Inclusive production}


An  interesting case is to compare the inclusive production for both equations. The inclusive production can be calculated using $k_t$ factorization, this  was proven for the case of hadron-nucleus interaction in Ref.\cite{KTINC}.
\bea \label{MF4}
&&\frac{d \sigma}{d y \,d^2 p_{T}}\,\, = \\
&&~~~~~~~~\frac{2C_F}{\alpha_s (2\pi)^4}\,\frac{1}{p^2_T}\int d^2 \vec b \,d^2 \vec B \,d^2 \vec r_T\,e^{i \vec{p}_T\cdot \vec{r}_T}\,\,\nabla^2_T\,N_G\Lb y_1 = \ln(1/x_1); r_T; b \Rb\,\,\nabla^2_T\,N_G\Lb y_2 = \ln(1/x_2); r_T; |\vec b-\vec B| \Rb. \nn
\eea

In the saturation region $\nabla^2 N_G \,=\, \nabla^2(2 N - N^2) \,\, \propto\,\, \exp\Lb- 2 \phi(z\Rb)$ and, therefore,
\beq \label{S17}
\frac{d \sigma}{d y \,d^2 p_{T}}\,\,\propto \,\,e^{ - 4 \phi(z)}
\eeq
 Using \fig{nz}-a we can  see that the effect  is large, and practically for $z > 3$ the ratio of  $exp\Lb  -4 \phi \Rb$
is negligibly small. In other words, the production of the gluons with transverse momenta smaller than $Q_s\exp[-1.5)$ is small.

 To estimate the inclusive production we need to know  $\nabla^2 N_G$. It is easy to see that
$\nabla^2 N_G\,\,=\,\,(1/r^2)d^2 N_G/d z^2$.
\eq{S8}, \eq{S8}, \eq{S15} and \eq{S16} allow us to calculate $\phi'_z $ and $\phi''_{zz}$.

~

~

~

Indeed,
\bea
\frac{d  \phi^{(0)}(z)}{d z} \,\,&=&\,\,\frac{1}{\sqrt{2}}\,\sqrt{\phi^{(0)}(z) + e^{- \phi^{(0)}(z)} - 1};\label{INC1}\\
\frac{d^2 \phi^{(0)}(z)}{d z^2} \,\, &=&\,\,\frac{1}{4}\Lb 1\,\,-\,\,e^{- \phi^{(0)}(z)}\Rb;\label{INC2}\\
\frac{d  \phi^{(1)}(z)}{d z} \,\,&=&\,\,\frac{ \frac{(1 + \bas)^2}{4} \tilde{N}\Lb z\Rb - \bas \,\Lb 1 - e^{ - \phi^{(1)}(z)}\Rb^2}{1\,\,-\,\,\bas\,\tilde{N}\Lb z\Rb};\label{INC3}\\
\frac{d^2 \phi^{(1)}(z)}{d z^2} \,\, &=&\,\,\frac{\bas \Lb 1 - e^{- \phi^{(1)}(z)}\Rb\Big\{ \frac{(1 + \bas)^2}{4} \tilde{N}\Lb z\Rb - \bas \,\Lb 1 - e^{ - \phi^{(1)}(z)}\Rb^2\Big\}}{\Lb  1\,\,-\,\,\bas\tilde{N}\Lb z\Rb\Rb^2} \,\nn\\
 &+& \, \frac{\frac{(1 + \bas)^2}{4}  \,\Big( 1 \,-\,e^{ - \phi^{(1)}(z)}\Big) \,-\,2\,\bas \frac{d \phi^{(1)}(z)}{d z} \,e^{ - \phi^{(1)}\Lb z\Rb}\,\Lb 1 - e^{ - \phi^{(1)}(z)}\Rb}{
 1\,\,-\,\,\bas\tilde{N}\Lb z\Rb}; \label{INC41}
\eea

\FIGURE[h]{
\centerline{\epsfig{file=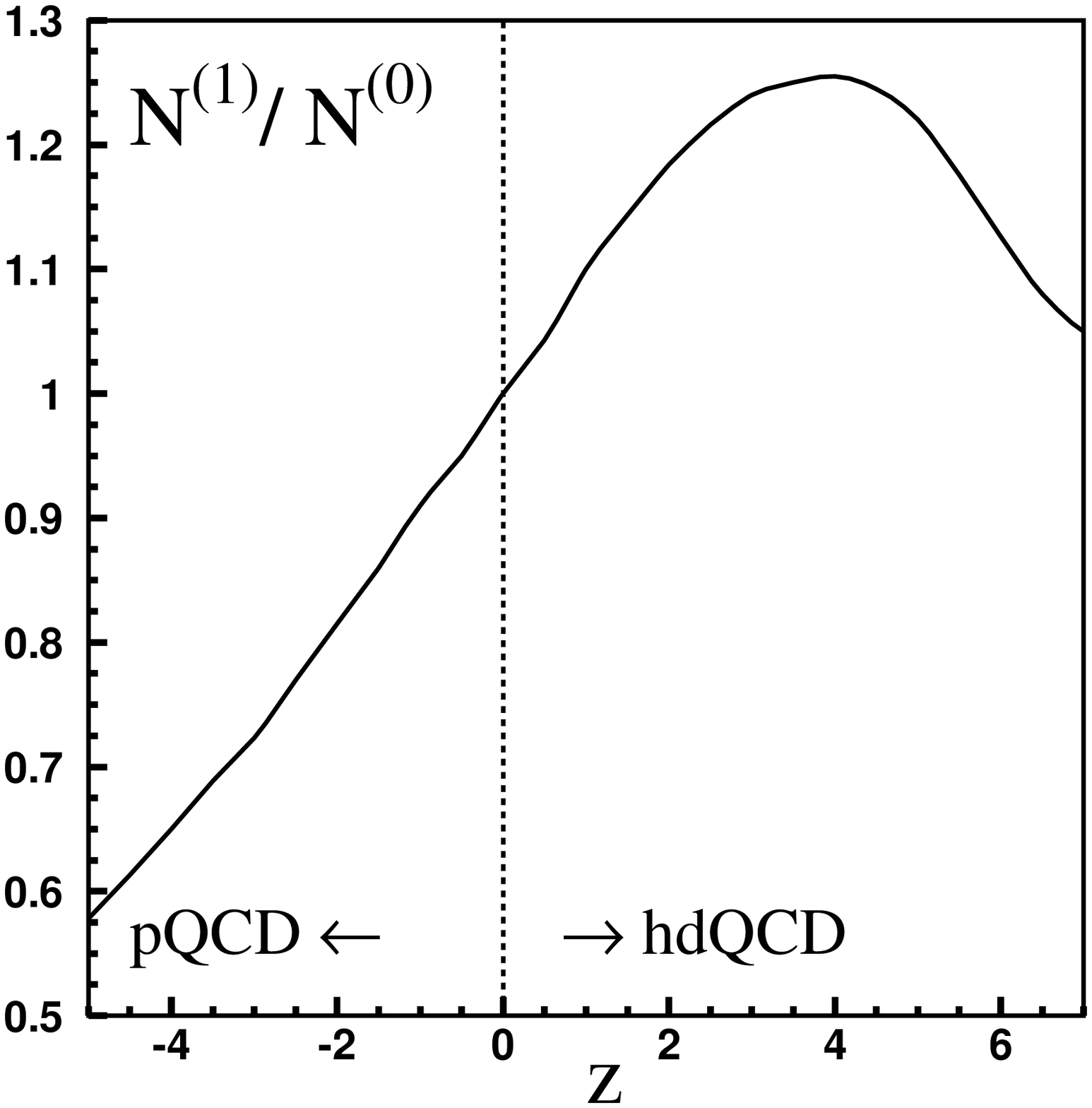,width=70mm}}
\caption{Dependence of the ratio $ N^{(1)}/N^{(0)}$  versus $z$ ($\bas = 0.2$,$\phi_0=0.15$, $ \tilde{N}_0 = 0.6$}
\label{nzr}}

In \fig{nz} the solutions for modified  equation and for the BK one are plotted. The comparison between them is more illustrative.


For $y_1=y_2$ \eq{MF4}  can be re-written in the form
\beq \label{MF5}
\frac{d \sigma}{d y \,d^2 p_{T}}\,\,= \,\frac{2C_F}{\as 2 (2\pi)^3}\,\frac{1}{x^2_\perp}\int d^2  b \,\,d^2  B \,\int^{+ \infty}_{- \infty}d  z \,e^{- z} \, J_0\Lb e^{\h z}\,x_\perp\Rb\,\,\frac{d^2N_G\Big( z ; b \Big)}{ d z^2}\,\,
\,\frac{d^2 N_G\Lb z ; |\vec b-\vec B| \Rb}{d z^2}.
\eeq

The integral over $z$ in \eq{MF5} includes the region for $z \,<\,0$, where the function $\phi(z)$ behaves as is shown in \eq{VICQS},
however  the kinematic region where we can trust \eq{VICQS} is rather narrow \cite{IIM}.  Using \eq{GALO} and \eq{NLO6} we can find the full solution for $\phi(z) $ at $z \,<\,0$:
\bea
\phi^{(0)}\Lb z; z \,<\,0\Rb &\,=\,& \phi_0\,\exp\Lb  \sqrt{ L\,\Lb L \,+\,z\Rb}\,-\,L\,-\,z\Rb; \label{INC21}\\
\phi^{(1)}\Lb z; z \,<\,0\Rb &\,=\,& \phi_0\,\exp\Lb \Lb 1 \,+\, \bas\Rb\,  \sqrt{ L\,\Lb L \,+\,z\Rb}\,-\,\Lb 1\,+\,\bas\Rb \,\Lb L\,+\,z\Rb\Rb; \label{INC22}
\eea
where $x_\perp \equiv p_T/Q_s\Lb Y\Rb$ and $L \,=\,4 \bas y$ for BK equation and $L = 4 \bas/(1 \,+\,\bas)^2\,y$ for the modified BK equation.

Using  these expressions we can calculate $d^2 N/d z^2$ for both cases at $z \to 0$ for $z < 0$:
\bea \label{INC4}
\frac{d^2 N^{(0)}}{d z^2} \xrightarrow{ z \to 0^-} \,\frac{\phi_0}{4}\,\Lb 1\,\,+\,\,\frac{1}{L} \Rb &  \neq & \frac{\phi_0}{4}\,\xleftarrow{z \to 0^+} \,\frac{d^2 N^{(0)}}{d z^2} \label{41}\\
\frac{d^2 N^{(1)}}{d z^2} \xrightarrow{ z \to 0^-} \,\frac{\phi_0\,\Lb 1 \,+\,\bas\Rb}{4}\,\Lb 1\,\,+\,\,\frac{1}{L}\Rb & \neq & \frac{\phi_0\,\Lb 1 \,+\,\bas\Rb}{4}\,\xleftarrow{z \to 0^+} \,\frac{d^2 N^{(1)}}{d z^2} \label{42}
\eea
From \eq{41} and \eq{42} we have that the double derivatives at $z=0$ only  match at large values of $L$. The calculations for the ratio
\beq \label{INC5}
R=\frac{d \sigma^{(1)}}{d y \,d^2 p_{T}} /\frac{ d \sigma^{(0)}}{d y \,d^2 p_{T}}
\eeq
are shown in \fig{incra}. One can see that the energy conservation could lead to the suppression of the inclusive production by a factor of two.

\subsection{New scale for freeze-out}
The most striking result of the modified BK equation is the appearance a new scale. Indeed, \eq{S13} and \eq{S14} show that
for $z$ when $\bas \tilde{N} \,\to\,1$ , the value of $\phi$ increases. Formally speaking for $\bas \tilde{N} \,>\,1$ $\phi$ becomes an  oscillating function which gives negligible contribution to the inclusive cross section (see \fig{ns}) .  Since in vicinity $\bas   \tilde{N} \,\approx\, 1$ the value of $\phi \propto 1/\bas^2 \,\geq \,1$, we can conclude from \eq{S1}, that $\tilde{N} \,=\,z$.
Therefore, the equation  $\bas \tilde{N} \,=\,1$ leads to $z = 1/\bas$, which gives a new scale
\beq \label{FR1}
Q_{fr}\,\,=\,\,Q_s\,\exp\Big( - \frac{1}{2 \bas}\Big)
\eeq

The physical meaning is clear: the dipoles with the size larger than $2/Q_{fr}$ will not produced. In other words,  in the  collisions we produce dipoles with the size $2/Q_{fr}$ with the typical temperature $T = Q_{fr}$.

\FIGURE[h]{
\centerline{\epsfig{file=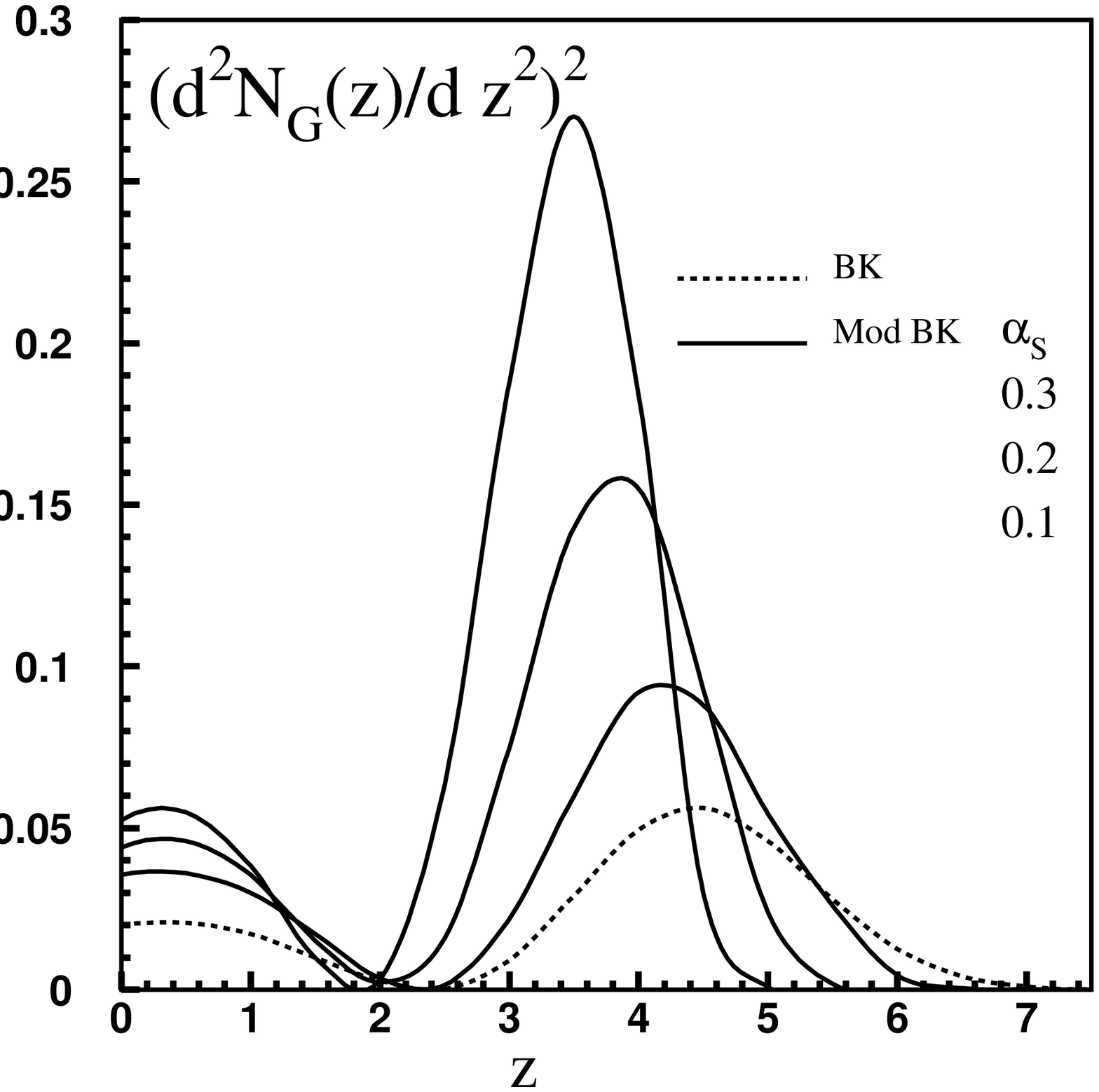,width=90mm}}
\caption{Dependence of $d^2 N/d z^2$ versus $z$ for the solution to BK equation (dashed line) and to modified BK equation at different values of $\as$.}
\label{ns}}

It should be stressed that we reconstruct the form of the non-linear term in the modified BK equation assuming the probability interpretation of the parton cascade, which was proven for the LO BK, however we believe on physical grounds   that it will also be
correct in the NLO. However, the NLO corrections also lead to the transition one dipole decays into three dipoles and induced by this transition term in the linear part of the equation.  We neglected these terms since they induce the $N^3$ term in the equation. This term is small in the vicinity of the critical line ( $N \propto \bas $ on it \cite{LT}) and only this kinematic region contributes to inclusive production.

The failure to obtain  large corrections from energy conservation forces us to believe  that non-perturbative contribution
will be the reason for the small value of the NMF( nuclear modification factor). It should be stressed that the dipole rescatterings has been taken into account in \eq{MF4}. A separate question is, could we trust the $k_t$ factorization formula of \eq{MF4} in the case of nucleus-nucleus collisions?  The rigorous answer is no since there is no proof of the $k_t$ factorization for this case. However, it is difficult to believe that the violation of the $k_t$ factorization will be responsible for the value of  NMF, Indeed, in \eq{MF4} and \eq{MF5} the main contribution stems from the vicinity of the saturation scale (the largest one in nucleus-nucleus scattering)  and in the case of  the two scale problem: high $p_T$ and the saturation scale,   $k_t$ factorization has been proven in Ref.\cite{KTINC}.

We think that the most interesting outcome of this  solution  is the appearance of the new scale $ Q_{fr} \,=\,Q_s \exp(- 1/(2 \bas))$  which characterize the temperature of the produced hadron ( $T\, \approx\,Q_{fr}$) . The production of dipoles with the size larger than  $2/Q_{fr}$ are suppressed. Much more work is need to answer the question whether this scale is an artifact of our simplification or  a general feature of the energy conservation.

\section{Impact parameter dependence}

In this section we address the problem of the  impact parameter ($b$)  dependence for both BK equation and modified BK equation. It is well known that $b$ dependence is one of the most challenging questions in perturbative QCD\cite{KOWI}. Indeed, in QCD we have the massless gluon,  and because of this at large values of $b$ we are doomed to have   power -like decrease of the scattering amplitude in both linear and non-linear equation.  Such a  decrease results in power-like increase of the radius of interaction $R \propto s^{\lambda}$, which indicates that the large values of $b$  are dominant in   the  equation, which  has been derived in perturbative QCD\cite{KOWI}. Therefore, at first sight the approach based on non-linear equation appears to be inconsistent. Unfortunately, as far as we know there exist only two papers (see Ref.\cite{BDEP}) where the numerical solutions were found  to BK equation including  $b$
dependence. These solutions confirm the result  that large values of $b$ become essential and some modifications of the approach based on the BK equation, are  required.

In such a situation the analytical solution that includes $b$ dependence even for the BK equation with simplified kernel, can be  instructive.
\subsection{Solution with the impact parameter dependence}
For simplified kernels of \eq{GALO} and \eq{NLO6} the $b$-dependence can be easily found following Ref.\cite{LT}. Indeed, in both equations the kernel for $r^2 Q^2_s < 1$ coincides with the DGLAP kernel.  In DGLAP evolution we sum the log of transverse momenta of the partons. In this integral the momentum transfer $q$ enters only as the low limit of integration over transverse momentum (say $k_\perp$). The integral has the following form
\beq
\label{BD1}
\ln \Lb Q^2/q^2\Rb \,\,=\,\,\int^Q_q\,\frac{d^2 k_\perp}{k^2_\perp}
\eeq

\FIGURE[h]{
\centerline{\epsfig{file=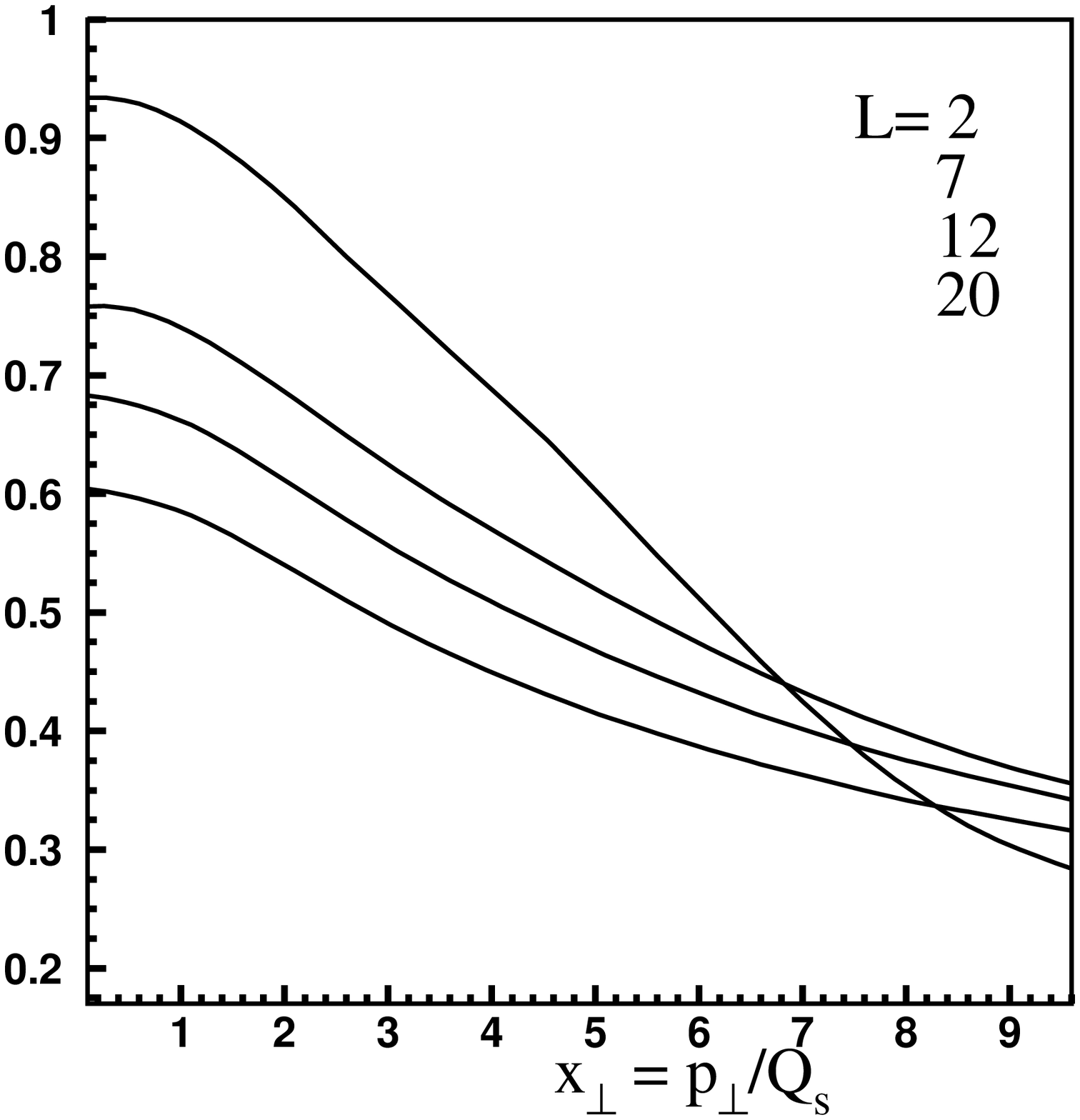,width=80mm}}
\caption{Dependence of the ratio $R=\frac{d \sigma^{(1)}}{d y \,d^2 p_{T}} /\frac{ d \sigma^{(0)}}{d y \,d^2 p_{T}} $ for $y_1 = y_2$  versus $x_\perp = p_{T}/Q_s$
and different values of  $L$ where $L\, =\, 4 \,\bas Y $ for the solution of BK equation and $L \,=\,  4 \,\bas/(1 + \bas)^2\,Y $ for the solution of the modified BK equation. In the picture the value of $L$ for the BK equation is shown. ($\bas = 0.2$, $\phi_0=0.15$,  $ \tilde{N}_0 = 0.6$)}
\label{incra}}

However, the low limit in \eq{BD1} is $q$ only in the region where $q > Q_0$ where $Q_0$  is the scale from with which we can apply perturbative QCD. Therefore, in the kinematic region where $q < Q_0$ no dependence of $q$ appears in the QCD evolution.
However, the non-perturbative corrections generate the non-perturbative form factor, and the general form of the solution for say, gluon density,  has the following form
\beq \label{BD2}
N\Lb z<0; r,  b; Y\Rb\,\,\,=\,\,S(b)\,N\Lb z<0; r ; Y\Rb\,\,\,\xrightarrow{z\,\to\,0}\,\,S\Lb b \Rb\,e^{ \Lb 1 - \gamma_{cr}\Rb\,z}
\eeq
In the vicinity of the saturation scale \eq{BD2} together with \eq{VICQS} leads to
\beq \label{BD3}
N\Lb z \,\xrightarrow{z<0}\,0 ; r,  b; Y\Rb\,\,\,=\,\,\Big( r^2\, Q^2_s(Y)\Big)^{ 1 \,-\,\gamma_{cr}}\,S\Lb b\Rb\,\,=\,\,
\Big( r^2\, Q^2_s(Y;b)\Big)^{ 1 \,-\,\gamma_{cr}}\,
\eeq
where
\beq \label{BD4}
Q^2_s\Lb Y;b\Rb\,\,=\,\,Q^2_s\Lb Y\Rb
\,\,S^{\frac{1}{1 - \ga_{cr}}}\Lb b \Rb
\eeq

The kernel of the BK equation as well as the modified BK equation does not depend on $b$. Therefore, the only information about $b$-dependence stems from the $b$-dependence on the critical line which can be absorbed in the dependence of the saturation scale on $b$ as  is given by \eq{BD4}. Finally, the solution of the equation will be the same as without $b$-dependence but with new variable
\beq
\tilde{z}\Lb z ;b\Rb\,\,\,=\,\,\ln  \Big( r^2\, Q^2_s(Y;b )\Big)\,\,=\,\,z \,\,+\,\,\frac{1}{1\,-\,\ga_{cr}}\, \ln\Lb S\Lb b \Rb\Rb
\eeq

This new variable alters the estimates for the new scale for freeze-out as one can see from \fig{ntil}, that shows $\tilde{N}$ as function of $z$ at different values of $b$. One can see that the value of $z$ at which $\tilde{N} \to 1/\bas$ is different for different $b$. In terms of ion-ion collisions this dependence can be translated to dependence on
the centrality cuts leading to temperature dependence for different centrality.

\FIGURE[h]{
\centerline{\epsfig{file=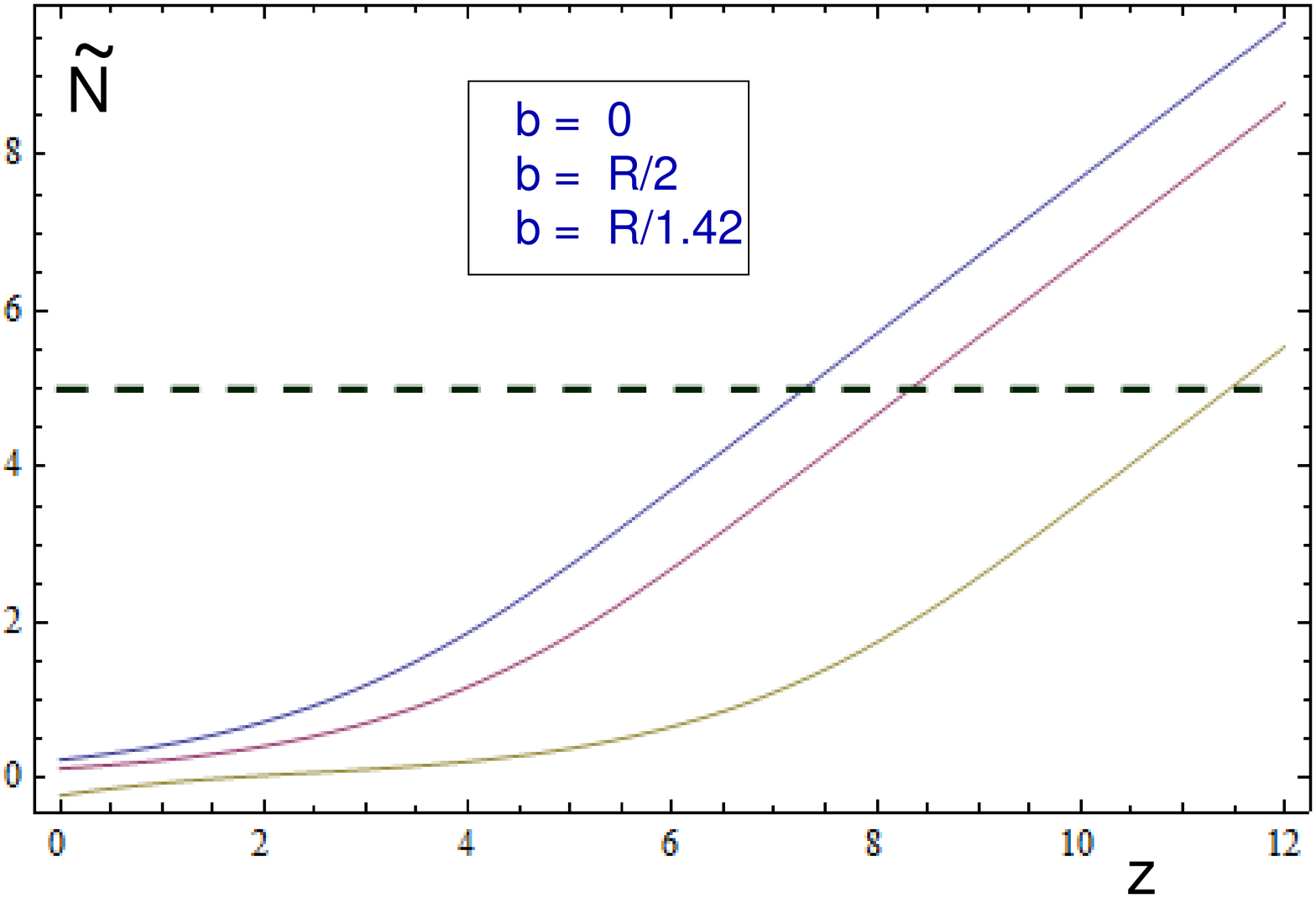,width=100mm}}
\caption{ $\tilde{N}$ versus $z$  and for solution to modified BK equation at different values of  $b$ ( see \eq{BD6}). The dotted line shows $\tilde{N} = 1/\bas $ for $\bas = 0.2$}
\label{ntil}}
\subsection{Inclusive production}
\eq{MF5} can be re-written in the form (for $y_1=y_2$)
\beq \label{BD5}
\frac{d \sigma}{d y \,d^2 p_{T}}\,\,= \,\frac{2C_F}{\as 2 (2\pi)^3}\,\frac{1}{x^2_\perp}\,,\int^{+ \infty}_{- \infty}d  z \,e^{- z} \, J_0\Lb e^{\h z}\,x_\perp\Rb\,\,\Big\{\int d^2 b \frac{d^2N_G\Big( \tilde{z}\Lb z; b\Rb \Big)}{ d z^2}\Big\}^2\,\,
\eeq
where $x_\perp$ is defined in \eq{MF5}.

As we have mentioned in section 3 we calculated the inclusive cross section assuming that $N \propto \Theta\Lb R \,-\,b\Rb$
Here, we choose the exponential dependence for $S\Lb b \Rb$
\beq \label{BD6}
S\Lb b \Rb\,\,\,=\,\,\,\exp\Lb - b^2/R^2\Rb\,\,
\eeq
We take $R^2 = 11\,GeV^{-2}$ for the nucleon and for nuclei we choose $R^2\,\,=\,\,(2/5)\,R^2_{WS}$ where $R_{WS}$ is the radius of the nucleus in Wood-Saxon parameterizations ($R_{WS} \,=\,6 \,A^{1/3}\,GeV^{-1}$).

Changing variable $b \to \rho = b/R$ oneobtains

\beq \label{BD8}
\frac{1}{\Lb \pi R^2\Rb^2}\,
\frac{d \sigma}{d y \,d^2 p_{T}}\,\,= \,\frac{2C_F}{\as 2 (2\pi)^3}\,\frac{1}{x^2_\perp}\,,\int^{+ \infty}_{- \infty}d  z \,e^{- z} \, J_0\Lb e^{\h z}\,x_\perp\Rb\,\,\Big\{\int d \rho^2 \frac{d^2N_G\Big( \tilde{z}\Lb z; \rho\Rb \Big)}{ d z^2}\Big\}^2\,\,=\,\,{\it  T}\Lb x_\perp\Rb
\eeq
Function ${\it T}\Lb x_\perp\Rb$ does not depend on the target. Therefore,\eq{BD8} shows that for the scattering of two
identical nuclei at $y_1 = y_2$ we expect  the scaling behavior: the inclusive production depends only on variable $x_\perp$, instead of
dependence on the number of nucleons ($A$),  rapidity $y=y_1=y_2$ and $p_T$.

\FIGURE[h]{
\centerline{\epsfig{file=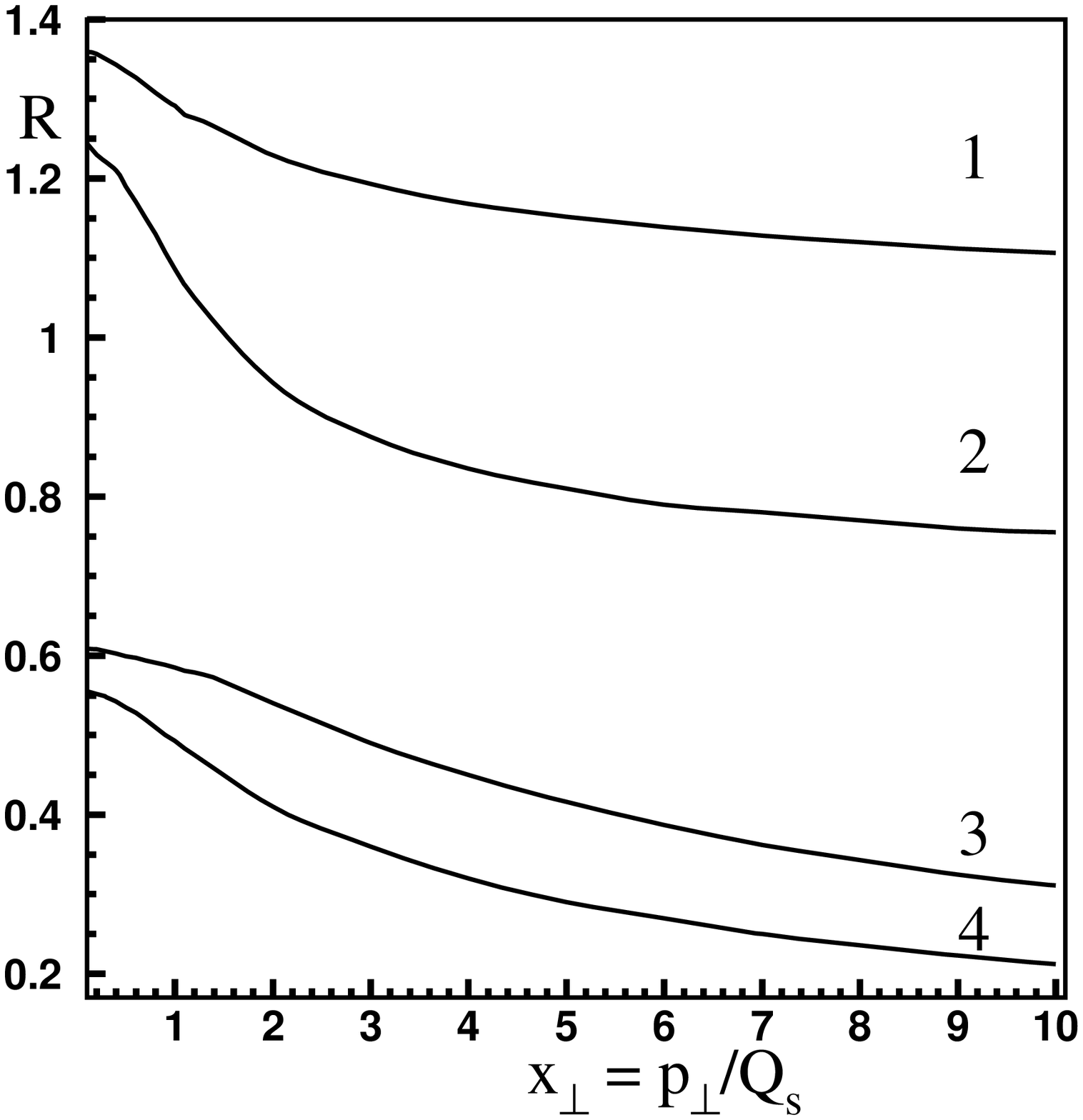,width=90mm}}
\caption{The ratio   $R\Lb b \Rb$ ( see \protect\eq{BD7})  versus $x_\perp$ for  the solution of the modified BK equation (curve 1) and for the BK equation (curve 2).  Curve 3 gives the ratio of $T(x_\perp)$ for BK equation to the modified BK equation without $b$-dependence while curve 4 gives the same ratio taking into account $b$-dependence.}
\label{rab}
}
\fig{rab} shows the influence of $b$ dependence of the inclusive production.

In \fig{rab} the ratio
\beq   \label{BD7}
R\Lb b \Rb \,\,\,=\,\,\,{\it T}\Big(\mbox{with $ b$ dependence}\Big)\Big{/} {\it T}\Big(\mbox{without
$ b$ dependence}\Big)
\eeq
is plotted.

One can see that that the solution of the BK equation with $b$ dependence,  differs from the solution of  the same equation without taking into account the $b$-dependence,  by  + 25\%  at small values of $x_\perp$ and and by -25\% at large values of $x_\perp$.
It is  surprizing that the solution of the modified BK equation leads to the same inclusive production at large $x_\perp$
both for the cases with $b$ and without $b$ dependence. However, at small $x_\perp$ the solution with $b$ dependence exceed the one without $b$-dependence by 35\%.  The differences between solutions to modified BK and BK equations are shown in curves 3 and 4 in \fig{rab}. The  solutions with $b$-dependence  could differ by a factor of  three at large values of $x_\perp$.

$T\Lb x_\perp\Rb $ as a function of $x_\perp$ is plotted in \fig{rat}. Note that the nuclear modification factor(NMF) can be easily re-written through $T \Lb x_\perp \Rb$, namely,
\beq \label{NMF}
\mbox{NMF}\,\,\equiv\,\,
\,\,\,\,\frac{1}{N_{coll}}\,
\,\frac{\frac{d^2 N_{AA }}{d y d^2 p_\perp}}{\frac{d^2 N_{NN}}{d y d^2 p_\perp}}\,\,=\,\frac{1}{A^{2/3}}\,\frac{{\it T} \Lb x_\perp\Rb}
{{\it T} \Lb x_\perp \frac{Q_{s,A}}{Q_{s,N}}\Rb}
\eeq
where $N_{coll}$ is the number of collisions. Using the  calculated $T \Lb x_\perp \Rb$ (see \fig{rat})  and the fact that $Q_{s,A}/Q_{s,N}\,=\,A^{1/6}$  we see that for the gold  NMF = (0.2 $\div$ 0.3) $\times $ 1.7 \,=\,0.3 $\div$ 0.5
for $x_\perp = 0.1 \div 10$ ( $p_\perp \,= 0.14 \div 14 \,GeV$) which is in about 2 times larger than the experimental NMF ( see  RHIC data \cite{BRAHMS0,BRAHMS1,BRAHMS2,PHENIX1}). However, the ratio of the radii ($R^4_A/R^4_N = 1.7$)  as well as Gaussian parametrization cannot be considered as reliable. Therefore, we can conclude that both BK equation and the modified BK equation lead to rather small NMF but perhaps, this suppression is not sufficient to describe the experimental data.

\FIGURE[h]{
\centerline{\epsfig{file=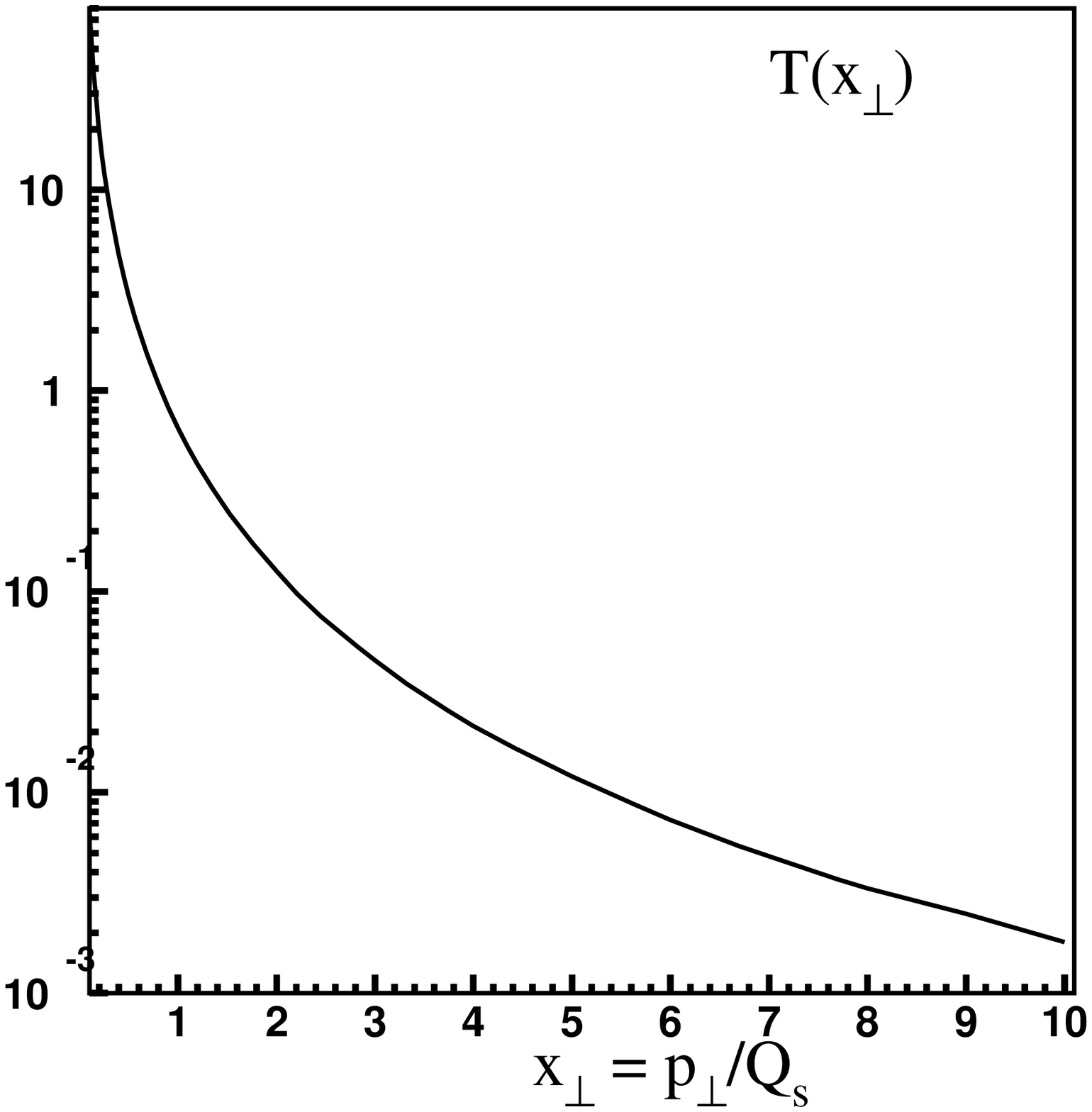,width=90mm}}
\caption{ $T\Lb x_\perp\Rb$ (see \eq{BD8}  versus $x_\perp = p_\perp/Q_s$ for the solution of BK equation that  takes into account $b$-dependence.}
\label{rat}
}

\section{Conclusions}

In this paper we study two questions: the influence of energy conservation on the solution to the non-linear equation; and
the $b$-dependence of the solution.
 The influence of energy conservation on the solution of the BK  equation  was investigated taking the impact parameter dependence in the form of $\Theta\Lb b - R\Rb$. The answer to this question is that  the energy conservation  reduces the value of the inclusive cross section by 20-40\%  at reasonable values of the QCD coupling ($\bas \approx 0.2$).  These estimates are based on the modified BK equation (see \eq{MODBK}) and on the form of the kernel given in \eq{NLO6}, which were derived from the form of the NLO corrections to the BFKL kernel, given in \eq{NLOOUR}.   The derivation of \eq{NLOOUR} stems from two observations: (i) the fact that \eq{ADEKL} describes the exact value of the anomalous dimension of the DGLAP evolution with a very good accuracy; and (ii) the suppression of anti-DGLAP evolution of the transverse momenta of partons in the non-linear equation (see section 2.1 of this paper).

It should be stressed that we reconstruct the form of the non-linear term in the modified BK equation assuming the probability interpretation of the parton cascade which was proven for the LO BK  but has so clear physical meaning   that we believe that it will be
correct in the NLO. However, the NLO corrections also lead to the transition of  one dipole decaying into three dipoles, and induced by this transition term in the linear part of the equation.  We neglected these terms since they induce the $N^3$ term in the equation. This term is small in the vicinity of the critical line ( $N \propto \bas $ on it \cite{LT}), and only this kinematic region contribute to inclusive production.

The failure to obtain the large corrections from the energy conservation suggests  that  the non-perturbative contribution
is  the reason for the small value of the NMF( nuclear modification factor). We wish to  stress that the dipole rescatterings has been taken into account in \eq{MF4}. The separate question is can  we trust the $k_t$ factorization formula of \eq{MF4} in the case of nucleus-nucleus collisions?  The rigorous answer is no,  since there is no proof of the $k_t$ factorization for this case. However, it is difficult to believe that the violation of the $k_t$ factorization will be responsible for the value of  NMF.  Indeed, in \eq{MF4} and \eq{MF5} the main contribution stems from the vicinity of the saturation scale (the largest one in nucleus-nucleus scattering)  and in the case of  two scale problems: high $p_T$ and the saturation scale the $k_t$ factorization has been proven in Ref.\cite{KTINC}.

In our opinion  the most interesting result of this solution  is the appearance of the new scale   $ Q_{fr} \,=\,Q_s \exp(- 1/(2 \bas))$, which characterize the temperature of the produced hadron ( $T\, \approx\,Q_{fr}$) . The production of dipoles with the sizes larger than    $2/Q_{fr}$ are suppressed. Much more work is needed to answer the question whether this scale is the artifact of our simplification or a general feature of  energy conservation.

As far as $b$ dependence is concerned, we found that the BK equation without $b$-dependence cannot guarantee an accuracy of calculation better than $\pm 25\%$. We consider an important observation that in our simplified equation  the entire $b$-dependence can be absorbed in the non-perturbative behavior of the saturation scale. We hope that this fact  can lead  to
a general scheme for taking into account the impact parameter dependence in the framework of high density QCD, however 
to  rewrite  the non-linear equation that absorbs the new non-perturbative scale, still  needs more work.

\section*{Acknowledgments}
We thank  Asher Gotsman and Jeremy Miller for fruitful discussions on the subject and for reading our manuscript.
 This work was supported in part by the  Fondecyt (Chile) grant  \# 1100648.


\end{document}